\documentclass[12pt]{article}
\usepackage{epsfig}
\begin {document}
\hoffset=-1.4cm
\voffset=1.0cm
\newcommand{\siganti}{\sigma^{\leftarrow \Rightarrow}}
\newcommand{\sigpar}{\sigma^{\leftarrow \Leftarrow}}
\begin{titlepage}

\hspace*{11cm} BONN-HE-2002-03

\vspace*{2.5cm}
\begin{center}
{\LARGE {\bf An Introduction to 
the Evaluation of \\
 Spin Structure Functions \\
\vspace*{2mm}
 from         Experimental Data}} 
\vspace*{1cm}
\normalsize

R.Windmolders \footnote{Supported by the Bundesministerium f\"ur Bildung und Forschung, contract Nr 06BN908I}.     \\
{\it Physikalisches Institut Universit\"{a}t Bonn, Nussallee 12 D-53115 Bonn, Germany}
 \vspace*{1cm}
\\
\vspace{2.0cm}
\today
\end{center}
\begin{abstract}
     These lectures  introduce the non-specialist to  the evaluation of  spin
structure functions  from asymmetries measured 
 in polarized deep-inelastic scattering experiments. The various steps leading from apparatus dependent
counting rate asymmetries to physics asymmetries are described. Special attention is given to
the effects of time variation in detector acceptances, to the use of deuterium as a neutron target and
to the corrections 
 due to the presence of unpolarized material in  a polarized target.
These topics are illustrated by examples taken from the CERN muon experiments.
\\

{\it Lectures given at the 10th "S\'eminaire Rhodanien de Physique" held at the Villa Gualino,
Torino, March 4-8, 2002.}
\end{abstract}
\end{titlepage}
\section { Introduction.}
\vspace*{0.2cm}

\begin{figure}[here]
\begin{center}
\epsfig{file=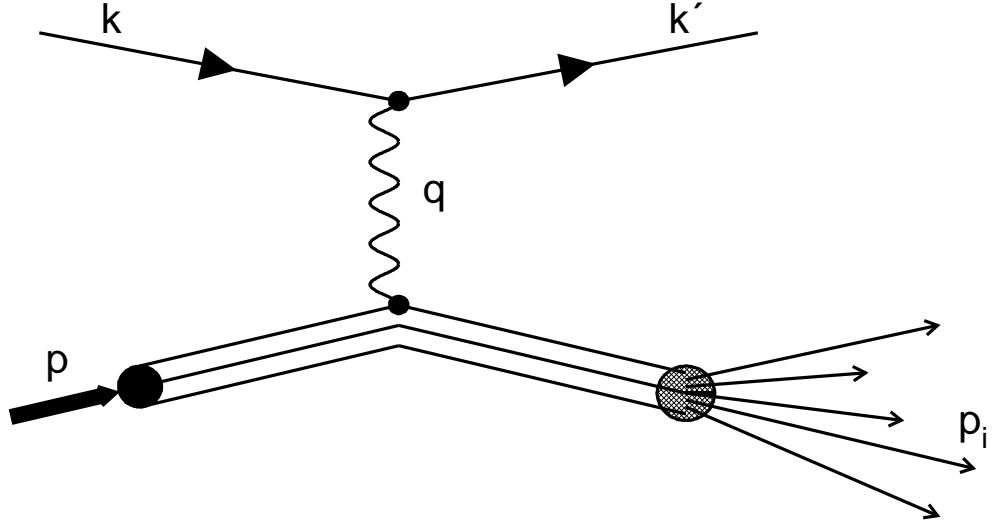,width=120mm}
\caption{\small 
\it  
Deep inelastic scattering of a lepton with four-momentum $k$ on a nucleon            
with four-momentum $p$ resulting into a lepton with four-mentum $k'$ and
a hadronic system.}
\label{fig:dis}
\end{center}
\end{figure}

In these lectures we discuss  experimental problems arising in the evaluation of the nucleon spin structure
from data taken in high energy polarized lepton-nucleon scattering experiments. \\
A general introduction to the formalism of nucleon spin structure functions as well as their interpretation
in terms of constituants can be found in recent textbooks on particle physics \cite{renton}. Only a short
overview will be presented here in order to define the kinematic variables and to introduce
the relevant physics parameters. 
Spin structure functions are measured in  experiments
performed with incident electron or muon beams. These experiments are called "inclusive" because only
the incident and scattered lepton are measured:
\begin{equation}
  \ell ~ N  \rightarrow  \ell^{'}   ~X.
\end{equation}
The symbol $X$ represents the unmeasured hadron final state which generally consists of
several particles. The kinematics of the reaction is entirely determined by  2 variables
$\nu$ and $Q^2$ which, in the lab system, are respectively the energy 
of the exchanged virtual photon 
and minus the 
square of its mass (Fig. \ref{fig:dis}):
\begin{equation}
\begin{array}{lcl}
\nu & =&  k_0 ~-~ k_0'   \\
Q^2 & =& -q^2 ~=~ ({\overline k } - {\overline k'})^2 - (k_0 - k_0')^2.
\end{array}
\end{equation}
In these lectures, we will consider only the region of deep inelastic scattering ("DIS"), where
the mass of the hadron system $X$ is much larger than the proton mass. This 
condition is equivalent to requiring sufficiently large values of $\nu$ and $Q^2$.
We also assume that the exchange of a virtual photon is the only process contributing
significantly to the reaction, which means that  weak interactions mediated by
the exchange of a $Z0$ or $W$ gauge boson are negligeable. In practice, the latter condition is satisfied
for all fixed target experiments.
\\

The mean feature of lepton scattering in the DIS region is the {\bf approximate scaling
of the differential cross sections}: for sufficiently large values of 
 $\nu$ and $Q^2$, the cross 
section only depends on the scaling variable defined by
\begin{equation}
x = \frac{Q^2}{2 M \nu}
\end{equation}
where $M$ is the nucleon mass. The discovery of scaling in 1967 has given
an experimental basis to the quark model. Scaling means that, at fixed $x$, the cross
section does not depend on the value of $Q^2$ or, in other words, on the size of the probed object.
Scaling therefore suggests that the scattering must take 
place on pointlike constituants ("partons") of the nucleon. In the quark-parton
model, the scaling variable $x$ is  equal to the fraction of the nucleon momentum
carried by  the quark which has absorbed the virtual photon.
\\

Spin effects generate a difference between the cross sections for parallel and antiparallel
orientations of the beam and target spins $\sigpar$ and $\siganti$. In order to discuss these
effects, we introduce the {\bf spin averaged cross section}
\begin{equation}
{\overline \sigma} = \frac{1}{2} (\siganti + \sigpar)
\end{equation}
which is measured in unpolarized experiments,
and the {\bf spin dependent cross section}
\begin{equation}
\Delta \sigma = (\siganti - \sigpar),
\end{equation}
which can only be measured in experiments where beam and target are both polarized.
In the deep inelastic region, the differential cross sections are expressed in terms of
two {\bf structure functions} which depend mainly on the scaling variable $x$ and, 
because scaling is only approximate, to a smaller extend
on $Q^2$:
\begin{equation}
\begin{array}{lcl}
d^2 {\overline \sigma}/(dx ~dQ^2) &=& a ~F_1(x,Q^2) ~+~ b ~F_2(x,Q^2)  \\
d^2 \Delta \sigma/(dx ~dQ^2) &=& c ~g_1(x,Q^2) ~+~ d ~g_2(x,Q^2).
\end{array}
\end{equation}
The unpolarized structure functions $F_1$ and $F_2$ are related in the quark-parton model
by the Callan-Gross relation
\begin{equation}
2 ~x ~ F_1(x) = F_2(x)
\end{equation}
and more generally by the relation \footnote{ The symbol $\cong$ is used in Eqns.(8-11) to warn
the reader that a kinematic factor close to 1.0 for $\nu^2 \gg Q^2$ has been neglected.}
\begin{equation}
F_1(x,Q^2) \cong  \frac{F_2(x,Q^2)}{2 ~x ~(1 + R(x,Q^2))}
\end{equation}
where $R(x,Q^2)$ is found to be small. The  $Q^2$ dependence of the structure
functions ("scaling violation") is due to interactions between nucleon constituants and is
successfully described in the context of quantum chromodynamics ("QCD").\\
The second  structure function $g_2(x,Q^2)$ has been found to be small.
For the longitudinal spin configuration which is discussed here, 
its contribution  is further 
suppressed  due to the small value of the coefficient
$d$ and can safely be neglected. 
The {\bf cross section asymmetry} defined by
\begin{equation}
A_{\|} = \frac{\Delta \sigma}{2 {\overline \sigma}} 
\end{equation}
can then be written as
\begin{equation}
A_{\|} \cong  D \frac{g_1(x,Q^2)}{F_1(x,Q^2)}
\end{equation}
where the  coefficient $D$ is directly calculable from the kinematic factors $a$, $b$ and $c$
of Eqn.(6). The evaluation of the asymmetry $A_{\|}$, which
is the main purpose of most experiments in polarized DIS, 
gives access to the spin structure function $g_1$ using the   spin averaged
functions $F_1$ or $F_2$ known from unpolarized experiments.
The remaining part of these lectures will describe the various steps involved in 
the derivation of $A_{\|}$ from experimental data with special emphasis on some
problems specific to muon experiments.  
\\

At the constituant level,
spin effects in DIS  can be intuitively understood  by the fact
that a quark having its spin projection along the reference axis (+OZ) can  absorb a
virtual photon which has its spin projection along (-OZ) and flip its spin, while no absorption can occur
when the two spins in the initial state are oriented in the same direction. \\
Defining $q_i^{+}(x)$ and $q_i^{-}(x)$ as the distributions of quarks of flavor $i$ with
spin along or opposite the nucleon spin, we see that the absorption cross section for
virtual photons with spin projection opposite to the nucleon spin ($\sigma_{1/2}$) will
be proportionnal to $q_i^{+}$ while the absorption cross section for virtual photons
with spin parallel to the nucleon spin ($\sigma_{3/2}$) will be proportionnal to $q_i^{-}(x)$.
The {\bf virtual photon asymmetry} is obtained by summing over the quark flavors $i$ and multiplying
each term
by the square of the quark charge expressed in units of the 
electron charge $(e_i^2=4/9 {\rm~or~} 1/9)$:
\begin{equation}
A_1 = \frac{\sigma_{1/2} - \sigma_{3/2}}{\sigma_{1/2} + \sigma_{3/2}} \cong 
\frac{\sum e_i^2 (q_i^{+}(x) - q_i^{-}(x))}{\sum e_i^2 (q_i^{+}(x) + q_i^{-}(x))}.
\end{equation}
The denominator in the previous expression shows the well known decomposition of $F_1$ in terms of
quark flavors. The numerator provides the corresponding decomposition of $g_1$ in terms of the
quark spin distributions $\Delta q_i(x) = q_i^{+}(x) - q_i^{-}(x)$. Comparing with the definition
of $A_{\|}$ given in Eqn.(10), we see that the factor $D$ can be considered as the depolarization of
the virtual photon. This factor depends mainly on the fraction $y = \nu / k_0$ of the beam energy taken away
by the virtual photon and is close to 1 for virtual photons carrying nearly the total energy of
the incoming leptons.

\section {Polarized DIS experiments.}
\vspace*{0.2cm}

\subsection {General characteristics of electron and muon experiments.}
\vspace*{0.1cm}

The first polarized DIS experiments have been performed at SLAC
in the early 80's \cite{baum}. Experiments using polarized
electron beams are presently running at JLAB (Virginia) in the energy range
of 2-5 GeV \cite{jlab}, at SLAC \cite{dis01} in the range of 10-50 GeV
and at DESY at 27 GeV \cite{hermes}. \\
These experiments use high intensity beams (e.g. $10^{12}$ electrons per second at SLAC)
and consequently need only relatively small targets to
reach a high 
statistical accuracy. They
can invert very frequently either the beam polarization (JLAB, SLAC) or the target polarization 
(DESY) 
and are therefore not very
sensitive to systematic effects due to changes in detector acceptance.                       
Their kinematic range is however limited due to the relatively
low incident energy. \\

\begin{figure}[here]
\begin{center}
\epsfig{file=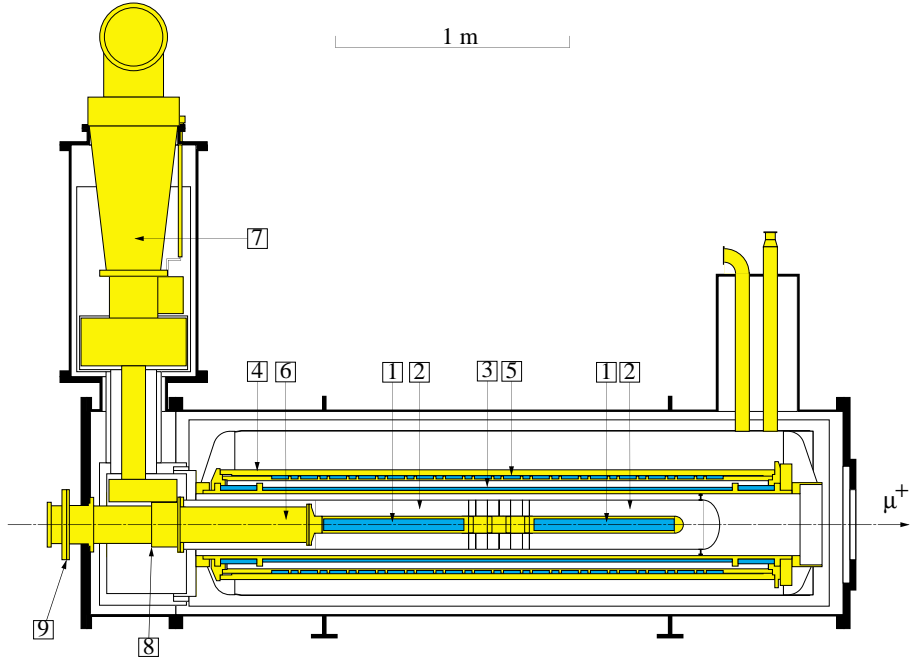,width=120mm}
\caption{\small
\it
The SMC target cryostat with the target holder as used since 1993. 
The muon beam traverses the cryostat from left to right. 
{\em (1)} target cells,
{\em (2)} microwave cavity, 
{\em (3)} solenoid coil, 
{\em (4)} dipole coil,
{\em (5)} correction coils, 
{\em (6)} dilution refrigerator, 
{\em (7)} precooler of $^3$He,
{\em (8)} indium seal, and 
{\em (9)} external seal.}
\label{fig:fcryo}
\end{center}
\end{figure}

The muon experiments, mainly performed at CERN, have  opposite characteristics: their
energy range is much higher (100-200 GeV) and their beam intensity much lower ($10^7$ muons per second,
i.e. down by a factor $10^5$ compared to SLAC). The incident muons are obtained from 2 body decays
of $\pi$'s and $K$'s and are naturally polarized due to the non-conservation of parity in the weak decay.
Positively charged muons
produced in a direction close to the primary hadron beam have a negative polarization of about 80 \% \cite{pbeam}.\\
Due to the limited intensity, muon experiments need large polarized targets.
Furthermore, in order to limit systematic effects
due to acceptance variation, it is essential to use simultaneously 2 target cells with
opposite polarization and to invert the polarization at regular intervals. The SMC target \cite{ptarg}
which has been in use at CERN since 1993 is shown as an exemple in Fig.~\ref{fig:fcryo} and will be
described in a further section. Since the time interval between 2 consecutive reversals of the
polarization is at least of the order of a few hours, the time dependence of detector acceptances 
becomes a critical issue in muon experiments. \\

\subsection {Counting rates and cross sections.}
\vspace*{0.1cm}

If we consider 2 consecutive runs with beam and target polarization parallel and antiparallel and with 
incident lepton fluxes $\phi _1$ and $\phi _2$ respectively, the numbers of collected interactions ($N_1,N_2$) will
be related
to the cross sections (4-5) by the following formulas:
\begin{equation}
\begin{array}{lcl}
N_1 &=& n ~\phi_1 ~a~ ({\overline \sigma} - (1/2) f ~P_b ~P_T ~\Delta \sigma)  \\
N_2 &=& n ~\phi_2 ~a~ ({\overline \sigma} + (1/2) f ~P_b ~P_T ~\Delta \sigma).
\end{array}
\end{equation}
Here $n$ stands for the area density of target nucleons, $a$ for the acceptance,
$f$ for the fraction of polarizable  material in the target and $P_b$ ($P_T$)
for the beam (target) polarization. At first sight, it looks straightforward
to extract $\Delta \sigma$ from these relations. However, one should
keep in mind that the spin dependent term is about 1000 times smaller than the
spin averaged one and that the  large fluxes cannot be determined with a precision
better than a few percent. In practice, the uncertainty on the fluxes totally masks the
spin contribution and makes a direct determination of $\Delta \sigma$
impossible. As a consequence, spin experiments do not measure cross sections but asymmetries.   
 
\subsection {Counting rate asymmetries  and spin asymmetries.}
\vspace*{0.1cm}

As an example we now consider 2 target cells of same density $n$ with 
opposite polarization exposed
to the same incident flux $\phi$.  \\
Under the unrealistic assumption that the spectrometer acceptance would be the same for the 2 cells,
the counting rate asymmetry 
\begin{equation}
\delta = \frac{N_2 - N_1}{N_2 + N_1}
\end{equation}
is obviously proportional to the cross section asymmetry defined by Eqn.(9):
\begin{equation}
\delta = f ~P_b ~P_T ~ A_{\|}.
\end{equation}
(We will see  in the
next section how the different acceptances for 2 target cells should be taken into account.) \\
The statistical error on $\delta$ 
\begin{equation}
\sigma (\delta) = \frac{(4 N_1 N_2)^{1/2}}{(N_1 + N_2)^{3/2}}
\end{equation}
reduces to $1/(N_1 + N_2)^{1/2}$ when the asymmetry is small. 
A large statistics will thus be required to reach a significant precision on a small
asymmetry. For instance, if $\delta \simeq 10^{-4}$, a relative precision 
$\sigma (\delta)/\delta = 0.1$ corresponds to a  $10^{10}$ interactions. \\
The statistical error on the physics asymmetry  $A_{\|}$ is further divided by the product $P_b~P_T~f$.
It is thus essential to keep these 3 factors as large as possible: a reduction 
by a factor $\alpha$ must indeed be compensated by increasing the number of interactions $N_{tot}$ by $\alpha^2$.
As a consequence, the {\bf figure of merit} 
for the comparison of  experiments performed with different beams
or different targets is given by  
\begin{equation}
(P_b ~ P_T ~ f)^2 ~N_{tot}.
\end{equation}

\begin{figure}[here]
\epsfxsize=8cm
\hfil
\epsffile[10 20 525 550]{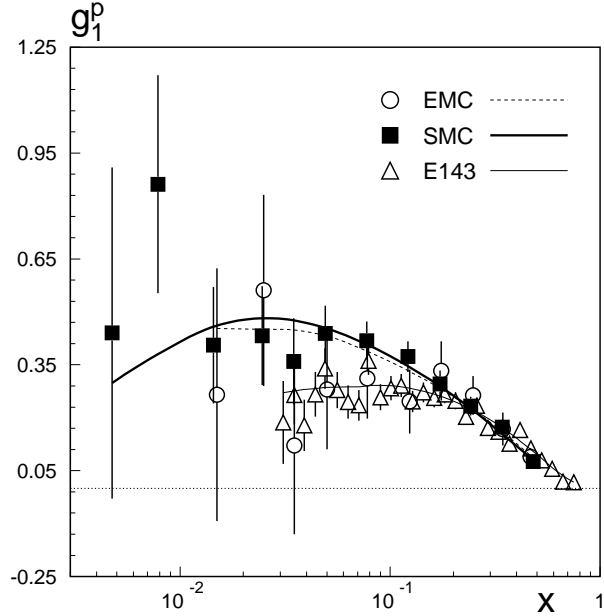}
\hfil
\caption{\small 
\it  Three data sets on $g_{1}^{\rm p}$. The curves show 
the QCD fit at the measured $Q^2$ of each data point. Error bars represent the total
error.}
\label{fig:qcdprot}
\end{figure}

All published values of the spin structure function $g_1$ have been obtained from cross section
asymmetries derived from counting rate asymmetries according to Eqn.(14).
The
relation between $A_{\|}$ and $g_1$
\begin{equation}
g_1(x,Q^2) = \frac{A_{\|}(x,Q^2)}{D} ~~\frac{F_2(x,Q^2)}{2 ~ x ~ (1 + R(x,Q^2)}
\end{equation}
involves the use of parametrizations of the unpolarized structure functions $F_2$ and
$R$ based on a large number of measurements from many experiments. The
uncertainty due to these parametrizations is included in the systematic error on $g_1$.
The statistical error on $g_1$ is propagated from the statistical error on $A_{\|}$ and
tends to increase at low $x$ due to the presence of the factor $x$ in the denominator.
This effect is well visible on the 3 data set shown in Fig.~\ref{fig:qcdprot}. The difference
between the $g_1$ values for experiment E143 with respect to the CERN experiments
reflects the $Q^2$ dependence of the structure function: at any fixed $x$, the average
$Q^2$ is about 6 times larger for the EMC-SMC data than for the E143 data. It can be observed
that this difference in $Q^2$ generates a sizable scaling violation in the region $x < 0.10$.
The curves on the figure show that this effect is well described in fits based on perturbative QCD \cite{qcd}. \\
 
\subsection {Acceptance corrected counting rate asymmetries.}
\vspace*{0.1cm}

We now come to the realistic case where two target 
cells (labelled "$u$" (up) and "$d$" (down)) 
are exposed to incident fluxes $\phi$ and $\phi '$ in 2 consecutive runs
\cite{smc_97}. In the first run, the "$u$" target polarization is opposite to the
beam polarization while the "$d$" target polarization is directed along the beam polarization. In the second
run, the polarizations of the 2 cells are reversed. Defining $m = P_b ~|P_T| ~f$, we obtain the 
following relations between numbers of events and cross sections:
\begin{equation}
\begin{array}{lcl}
N_u &=& n_u ~\phi ~a_u  ~{\overline \sigma} ~(1 ~+~ m ~A_{\|}),  \\
N_d &=& n_d ~\phi ~a_d  ~{\overline \sigma} ~(1 ~-~ m ~A_{\|}),  \\
N_u'&=& n_u ~\phi'~a_u' ~{\overline \sigma} ~(1 ~-~ m ~A_{\|}),   \\
N_d'&=& n_d ~\phi'~a_d' ~{\overline \sigma} ~(1 ~+~ m ~A_{\|}).
\end{array}
\end{equation}
The  counting rate asymmetries for the configurations before and after polarization reversal are
\begin{equation}
\nonumber
\delta ~=~ \frac{N_u - N_d}{N_u + N_d}, \\
\end{equation}
\begin{equation}
\nonumber
\delta '~=~ \frac{N_d' - N_u'}{N_d' + N_u'}.
\end{equation}
Eliminating the ratio $(n_d a_d)/(n_u a_u)$ between the 2 previous relations 
and defining the ratio of ratios of acceptance before and after reversal $K = (a'_d/a'_u)/(a_d/a_u)$, we obtain
\begin{equation}
\frac{(1 - mA_{\|})^2}{(1 + mA_{\|})^2} = K~ \frac{(1-\delta)(1 - \delta')}{(1 + \delta)(1 + \delta ')}
\end{equation}
which is a second order equation in $A_{\|}$. Keeping the solution satisfying $|A_{\|}| \le 1$ and 
approximating 
\begin{equation}
\Bigl ( (1 - \delta^2) (1 - \delta '^2) \Bigr )^{1/2}  \simeq 1 - (1/2)~ (\delta^2 + \delta'^2)
\end{equation}
we obtain for the case where $K = 1$:
\begin{equation}
A_{\|} = \frac {\delta + \delta '}{2 m}.
\end{equation}
Under the condition that the ratio of acceptances for the 2 target cells remains unchanged
after polarization reversal (i.e. $K = 1$), we thus obtain an unbiased estimate of the physics asymmetry
by taking the arithmetic mean of the  counting rate asymmetries before and after polarization reversal. \\

The derivation is slightly more complicated when $K$ is different from 1 \cite{caputo}.
In this case, since $K$ is nevertheless
close to 1, one may expand ${\sqrt K} = {\sqrt {1 + \epsilon}} \simeq 1 + \epsilon/2
- \epsilon^2/8$ and drop all square terms multiplied by $\epsilon$ in Eqn.(21) which
leads to
\begin{equation}
A_{\|} = \frac{(\delta + \delta')^2 - \epsilon (\delta + \delta ') + \epsilon^2/4}
{m ( 2 (\delta + \delta') - \epsilon (1 - \delta -\delta '))}.
\end{equation}
This in turn can be approximated by
\begin{equation}
A_{\|} = \frac{1}{m} (\frac{\delta + \delta '}{2} - \frac{\epsilon}{4} ).
\end{equation}
Here the first term correspond to the result obtained for the case $K = 1$.
The second term is a {\bf false asymmetry due to the change of the acceptance ratio}. We also observe
that even a small deviation of $K$ from the nominal value of 1 generates a sizable false asymmetry: in the case
where $(\delta + \delta')/2 \simeq 0.01$, a value $K = 0.97$  results in a false asymmetry
of the same order as the physics asymmetry and spoils the data. It is thus essential to keep
track of all changes in acceptances which may affect the value of $K$ and to eliminate
data sets where the ratio of acceptances deviates significantly from its nominal value.
The risk of such deviations obviously
decreases when the frequency of polarization reversals increases. Electron experiments 
where the polarization can be inverted for each accelerator pulse are  practically
unaffected by this false asymmetry while muon experiments need to reduce the time between consecutive
polarization reversals to the minimum acceptable.
 
\section {Solid polarized targets.}
\vspace*{0.2cm}

Solid polarized targets, such as the one described in \cite{ptarg}, are based on the 
alignment of spins in a magnetic field $H$. In  paramagnetic materials, spins of free electrons
align with the magnetic field which, at equilibrium, gives rise to a polarization
\begin{equation}
P ~=~ \frac{n_{+} - n_{-}}{n_{+} + n_{-}} ~=~ \tanh(\frac{\mu_B ~ H}{k ~ T}),
\end{equation}
where $\mu_B$ is the Bohr magneton and $k$ the Boltzmann constant.
For a field of $2.5 T$ and a temperature of 0.5 $K$, this corresponds to a polarization
of nearly 100 \%.  \\
Similar considerations apply to the proton spin but result, with the same field
and temperature, into a polarization of only 0.5 \%. It is thus essential to apply a process transfering
the electron polarization to the protons in order to obtain an usable polarized target. \\

Dynamic
nuclear polarization ("DNP") fulfills this requirement and can be understood at the most basic level
by considering a system composed of a non-interacting proton and electron in a magnetic field.
The 4 energy levels $|s_e,s_p>$  will be characterized by different projections of
the electron spin (first label) or the proton spin (second label) on the direction of
the magnetic field:  $|-,+>$, $|-,->$,
$|+,+>$ and $|+,->$ corresponding respectively to the energy levels $E_a$, $E_b$, $E_c$ and $E_d$.
The difference $E_c - E_a$ for opposite orientations of the electron spin corresponds
to a frequency $\nu_e = 70$ Ghz while the difference for nucleon spin orientations $E_b - E_a$
corresponds to $\nu_N = 106$ Mhz. At equilibrium, the relative population 
 is thus about 50 \% for each of the 2 lowest levels while the 2 other ones are nearly empty. \\
If a radio frequency $\nu = \nu_e + \nu_N$ is applied, a fraction $\tilde{\epsilon}$ of the population of level $a$
will be moved  to level $d$. This will be followed by a spontaneous transition from
level $d$ to level $b$ due to the strong coupling of the electron spins with the field and result into a
proton polarization  $P_N = 2 ~ \tilde{\epsilon}$. 
The opposite effect would be obtained by applying a radio frequency $\nu = \nu_e - \nu_N$. \\

DNP can only take place when several conditions are fulfilled: presence of the right amount of paramagnetic
centers in the material, a field strong enough to reach the appropriate separation of energy
levels and a temperature low enough to keep the width of these levels small. Under optimal
conditions, values of the proton polarization comparable to the electron polarization can be reached.
\\

\begin{figure}[here]
\begin{center}
\epsfig{file=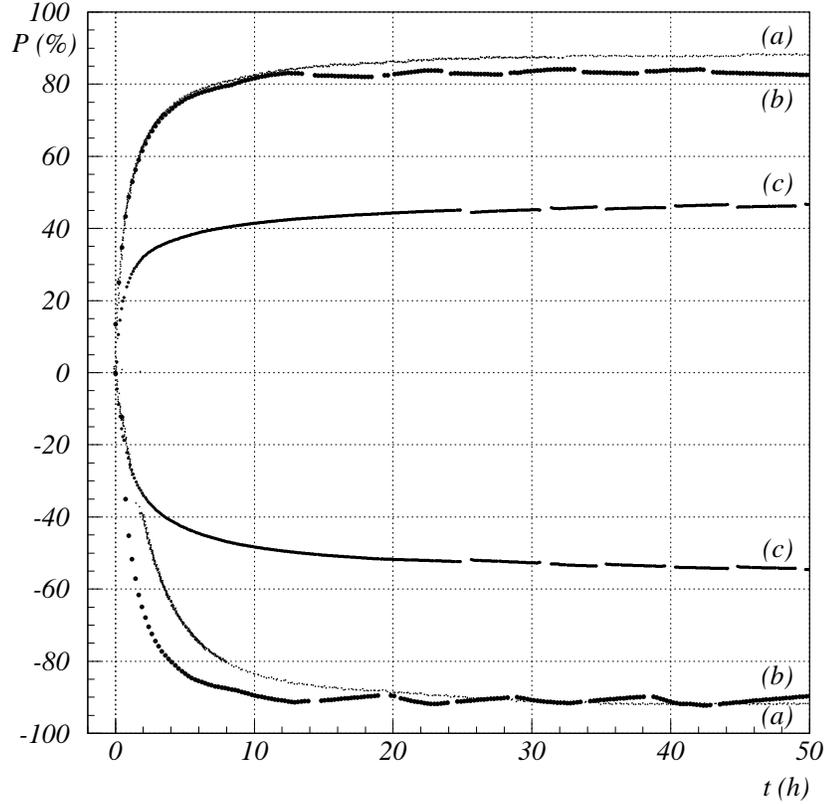,width=120mm}
\caption{\small
\it
Typical polarization build-up in the target materials, (a) ammonia, 
(b) butanol, and (c) deuterated butanol followed during 50 hours. 
The breaks in the data sets are interruptions of the measurements
due to field rotations.}
\label{fig:builtup}
\end{center}
\end{figure}

Fig.~\ref{fig:builtup} shows the polarization build-up for the various materials used in the
SMC experiment \cite{ptarg}. About 10 hours are needed before the polarization becomes close
to its maximum level. Changing the polarization by changing the radio-frequency is thus a
time consuming process which can only take place during a long interuption of the data taking.
More frequent polarization reversals are obtained by the so-called "field rotation":
the polarization is maintained by a transverse
field of 0.5 T produced by a dipole coil (Fig.~\ref{fig:fcryo}) while the solenoid field is 
ramped down and brought back in the opposite direction in a  process requiring less than half
an hour. 

\section {Deuterium as a neutron target.}
\vspace*{0.2cm}

The difference of the spin structure functions
$g_1$ for the proton and the neutron is one of the key issues in polarized DIS
\cite{qcd}. While proton data are obtained directly from hydrogen targets,
neutron data must be extracted from data taken on a deuterium or helium target. 
In this section, we will discuss several problems related to
the use of polarized deuterium as an effective neutron target.
 
\subsection {Spin asymmetries for a polarized deuteron.}
\vspace*{0.1cm}

Since the deuteron has spin one,  3  states with
spin projections +1, 0 and -1 on the reference axis have to be considered. The {\bf vector polarization} $P$
is defined as the difference between the number of deuterons with spin projections +1
and -1 by the total :
\begin{equation}
P = \frac{n_{+} - n_{-}}{n_{+} + n_{-} + n_{0}}.
\end{equation}
The number of interactions in 2 target cells ("$u$" and "$d$") with
polarization anti-parallel and parallel to the
beam polarization will be given by
\begin{equation}
\begin{array}{lcl}
N_u &=& \phi ~a_u ~(n_{+} \siganti + n_{-} \sigpar + n_{0} \sigma^{\leftarrow 0}) \\
N_d &=& \phi ~a_d ~(n_{-} \siganti + n_{+} \sigpar + n_{0} \sigma^{\leftarrow 0}).  
\end{array}
\end{equation}
For simplicity we assume here $a_u = a_d = 1$, so that the counting rate asymmetry
$\delta = (N_u - N_d)/(N_u + N_d)$ becomes
\begin{equation}
\delta = \frac{(n_{+} - n_{-})(\siganti - \sigpar)}{(n_{+} + n_{-})(\siganti + \sigpar) + 2 n_0 \sigma^{\leftarrow 0}}.
\end{equation}
 If in addition we assume that the {\bf tensor asymmetry} is zero, the 3 spin cross sections are related by
\begin{equation}
2 ~ \sigma^{\leftarrow 0} = \siganti + \sigpar
\end{equation}
and the counting rate asymmetry reduces to
\begin{equation}
\delta = P ~ \frac{\Delta \sigma}{ 2 {\overline \sigma}} = P ~ A_{\|},
\end{equation}
i.e. to the same relation as for spin 1/2 particles. The formalism developped in the
previous sections for the proton remains thus fully applicable to the deuteron, with
the only restriction that the polarization (Eqn.27) is defined in a different way. 
This different definition explains the lower values reached in the build-up of the
deuterated butanol  polarization as shown by curve (c) on Fig.~\ref{fig:builtup}.
 
\subsection {Deuteron spin and nucleon spin.}
\vspace*{0.1cm}

When the deuteron is in a S state, it is obvious that the proton and the neutron spin
are aligned with the deuteron spin, i.e. both nucleon spin projections are +1/2 when
the deuteron spin projection is +1. The situation is more complicated when the deuteron
is in D state, since the L = 2 angular momentum has to be combined with the nucleon
spins \cite{rondon}. As an example, the deuteron state $|J,J_Z> = |1,1>$ is obtained from the following
combination of orbital momentum $|L,L_Z>$ and total nucleon spin $|S,S_Z>$:
\begin{equation}
|1,1> = \sqrt{3/5} ~ |2,2> ~ |1,-1> - \sqrt{3/10} ~|2,1> ~|1,0> + \sqrt{1/10} ~|2,0> ~|1,1>.
\end{equation}
There is thus a probability of 3/5 to have both nucleon spins opposed
to the deuteron spin, a probability of 3/10 to have one of the nucleon spins aligned with
the deuteron spin and the other one in the opposite direction and a probability of 1/10
to have the 2 nucleon spins aligned with the deuteron spin. In total, when the deuteron is
in D state, the probability to have one of the nucleon spins opposed to the deuteron spin
is 0.75. The D state probability itself $\omega_d$ is of the order of 6 \% in all models
of the deuteron. \\
Defining an "average nucleon $N$" with $\sigma_N = (\sigma_p + \sigma_n)/2~$, we obtain
the following relation between the deuteron and nucleon spin cross sections:
\begin{equation}
\begin{array}{lcl}
\siganti_d &=& (1 - 0.75 ~ \omega_d)~  \siganti_N ~+~ 0.75~  \omega_d~ \sigpar_N  \\
\sigpar _d &=& (1 - 0.75 ~ \omega_d)~  \sigpar_N  ~+~ 0.75~  \omega_d~ \siganti_N .
\end{array}
\end{equation}
The difference between these 2 relations yields
\begin{equation}
g_1^d(x) = (1/2) ~ (1 - 1.5 ~ \omega_d) ~ (g_1^p(x) + g_1^n(x)).
\end{equation}
 
\subsection {Deuteron and nucleon
spin asymmetries.}
\vspace*{0.1cm}

\begin{figure}[here]
\begin{center}
\mbox{\epsfig{figure=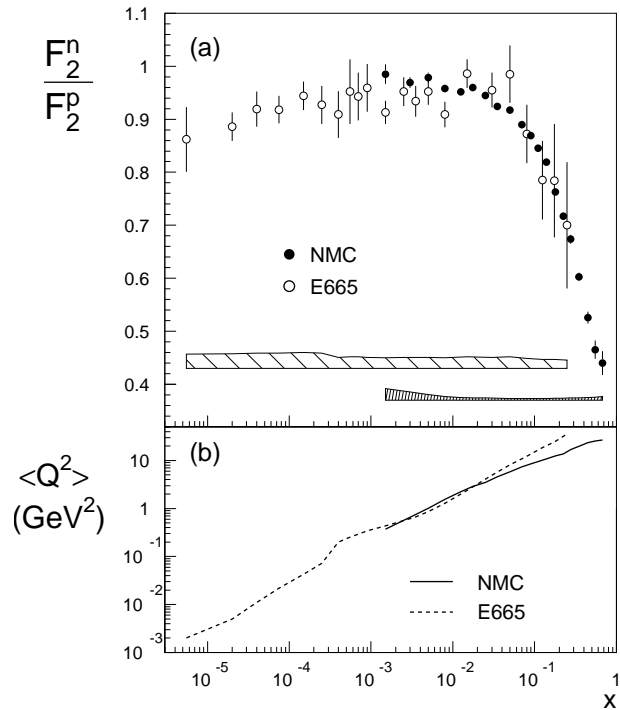,width=90mm}}
\end{center}
\caption{\small
\it
(a) Comparison of  the $x$ dependence of the NMC data for
$F_{2}^{\rm n}/F_{2}^{\rm p}$  with the results from the
experiment E665
~\protect\cite{e665}. The error bars correspond to the statistical errors,
the shaded bands to the systematic errors.
~~(b) Average $Q^2$ of the 2 data sets as a function of $x$.}
\label{fig:f2np_e665}
\end{figure}

The deuteron spin asymmetry
\begin{equation}
A^d_{\|} ~=~ (1 -1.5 ~\omega_d) ~\frac{\Delta \sigma_N}{2 {\overline \sigma_N}}
\end{equation}
can be split into a proton and a neutron term:
\begin{equation}
A^d_{\|} ~=~ (1 -1.5 ~\omega_d) \Bigl [~\frac{\Delta \sigma_p}{2 {\overline \sigma_p}}
~\frac{ {\overline \sigma_p}}{ {\overline \sigma_N}} +
~\frac{\Delta \sigma_n}{2 {\overline \sigma_n}}
~\frac{{\overline  \sigma_n}}{ {\overline \sigma_N}} \Bigr ].
\end{equation}
 Replacing the ratios of cross sections
by the corresponding ratios of $F_2$'s and taking out the virtual photon depolarization
factor one obtains:
\begin{equation}
A_1^d(x) ~=~ \frac{(1 - 1.5 ~\omega_d)}{1 ~+~ F_2^n(x)/F_2^p(x)} \Bigl( A_1^p(x) ~+~ \frac{F_2^n(x)}{F_2^p(x)} ~A_1^n(x) \Bigr ),
\end{equation}
which is the basic relation to extract neutron asymmetries from deuteron data and from  previously
measured proton asymmetries. The ratio $F_2^n(x)/F_2^p(x)$ has been measured with great precision in
several unpolarized experiments \cite{f2np,e665} and decreases rapidly for $x \ge 0.05$ (Fig.~\ref{fig:f2np_e665}).
 
As a consequence, deuteron data at larger $x$ give relatively little information about the neutron
asymmetry. It is only at very low $x$ ($\le 0.01$) that $A_1^d$ can be considered in first approximation as
the average of the proton and neutron asymmetries. \\
 
\section {Unpolarized material in polarized targets.}
\vspace*{0.2cm}

\begin{figure}[bottom]
\begin{center}
\mbox{\epsfig{figure=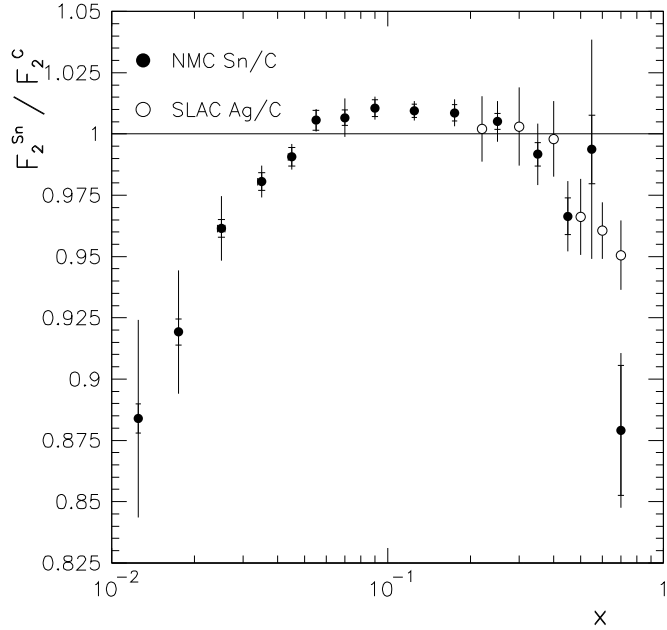,width=120mm}}
\end{center}
\caption{\small
\it
Structure function ratios for thin and carbon as a function of $x$, averaged over
$Q^2$ ~\protect\cite{f2ac}. The inner error bars represent the statistical uncertainty, the
outer errors the statistical and systematic uncertainties added in
quadrature.  The normalisation error is 0.2\% and is not included in
the systematic uncertainty.  The SLAC-E139~\protect\cite{slac_52} ratios
for silver and carbon are also plotted (open points).}
\label{fig:ratio}
\end{figure}

The fraction of "useful" (polarizable) material in a polarized target can be evaluated in
very first approximation by the ratio of the number of polarizable nucleons to the total number:
3/17 for ${\rm NH_3}$, 10/74 for butanol (${\rm C_4H_9OH}$), 20/84 for deuterated butanol (${\rm C_4D_9OD}$) and
4/8 for deuterated lithium (${\rm LiD}$) when ${\rm Li}$ is considered as (${\rm He^4} + {\rm D}$). However,a much more detailed 
evaluation is  needed for 2 reasons:
\begin{itemize}
\item {\bf all elements} of the target have to be taken into account, including impurities in
the target material, target windows, cooling mixture and coils used for the measurement of the polarization
if they are embedded in the material;
\item {\bf  the number of nucleons} in the ratios must be weighted by the respective cross sections.
\end{itemize}
We will only discuss here the second point, which has some general implications related to lepton-nucleon
interactions and will assume, for clarity, that we are dealing with a proton target ("$H$"). \\
In general, the dilution factor is defined as
\begin{equation}
f(x) = \frac{n_H ~ {\overline \sigma}_H^T}{\sum _A n_A ~ {\overline \sigma}_{N(A)}^T}
\end{equation}
where the index "$T$" refers to  total cross sections (including radiative effects).
In this formula, $n_A$ is the total number of nucleons (for the full target) in nuclei 
with atomic number $A$ and ${\overline \sigma}_{N(A)}$ the unpolarized cross section {\bf per nucleon}
on nucleus ${\rm A}$. \\
Weighting by cross sections is essential because the cross section on a nucleon bound in a nucleus differs
from the cross section on a free nucleon. The ratio 
\begin{equation}
{\overline \sigma_{N(C)}} / {\overline \sigma_{N}}  \simeq F_2^C(x,Q^2) / F_2^N(x,Q^2)
\end{equation}
has been measured with high precision over wide ranges of $x$ and $Q^2$ \cite{nmc_96}.
Extensive studies have also been performed 
for a large choice of other nuclei (for a review, see \cite{arneodo}). The ratios     
$F_2^A / F_2^C = (F_2^A/F_2^N)/(F_2^C/F_2^N)$ always
present a very characteristic dependence on $x$, as illustrated in Fig.~\ref{fig:ratio} for
Sn and Ag.

\begin{figure}[here]
\begin{center}
\mbox{\epsfig{figure=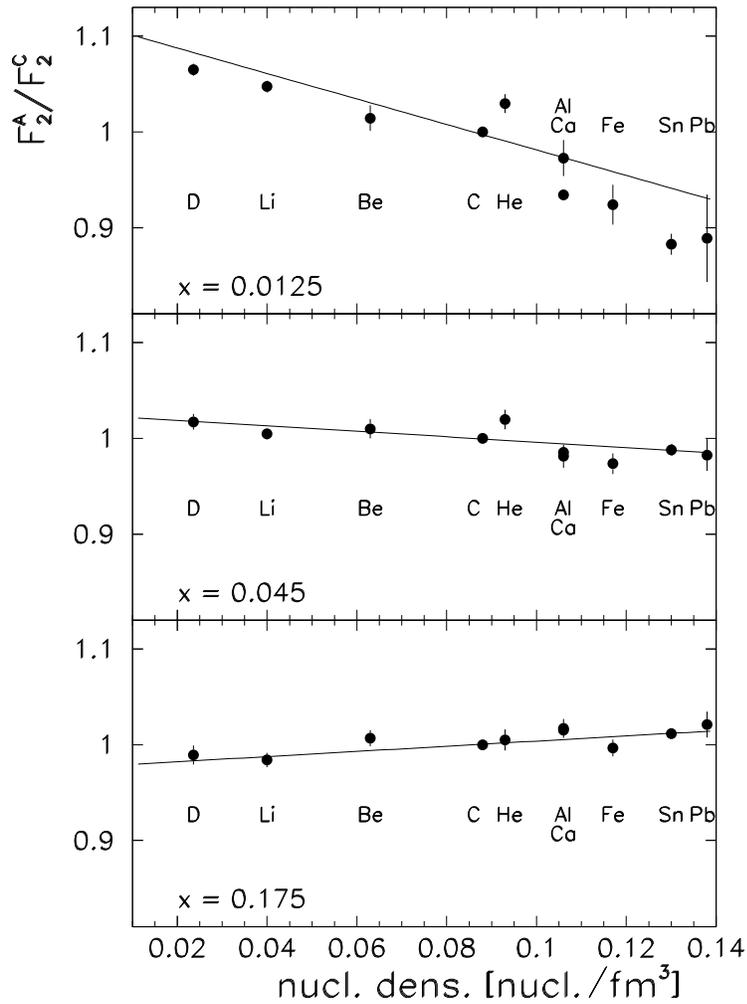,width=100mm}}
\end{center}
\caption{\small
\it
Structure function ratios measured by the NMC versus nuclear density $\rho$  at 
$x = 0.0125$, $x = 0.045$ and $x = 0.175$~\protect\cite{n_dens}. The solid lines 
show the result of a fit to the data with the function
$F_2^{\rm A}/F_2^{\rm C} =
\beta + \delta \rho(A)$. The errors shown are statistical 
only. 
}
\label{fig:dens}
\end{figure}

In the very small $x$ region, the cross section on a bound nucleon is reduced with respect to the
cross section on a free nucleon. This effect, known as "shadowing" becomes more pronounced for
heavier nuclei. Its name originates from a geometrical interpretation where  nucleons inside
the nucleus are assumed to be screened by those at the outer surface,  resulting in a total cross
section  proportionnal to ${\rm A}^{2/3}$ rather than to ${\rm A}$. At slightly larger
$x$ (0.05 $\le x \le$ 0.2), the cross section on a bound nucleon is  slightly larger  than on a free nucleon,
while, for $x \ge 0.3$, the ratio (39) drops again due to the so-called "EMC effect" \cite{emc}. \\
Measured cross section ratios with respect to ${\rm C}$ or ${\rm D}$ exist for a large number of nuclei.
For the remaining ones, the ratio at a given $x$  can be approximated by an interpolation
as a function of ${\rm A}$ as  shown in Fig.~\ref{fig:dens} \cite{n_dens}.\\

\begin{figure}[here]
\begin{center}
\mbox{\epsfig{figure=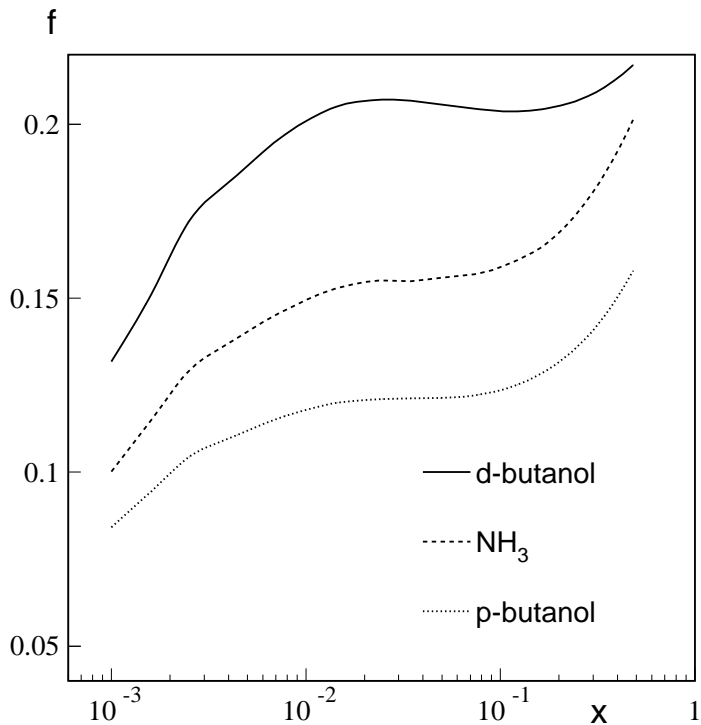,width=100mm}}
\end{center}
\caption{\small
\it
The dilution factors $f$ for the different target materials iused in the SMC experiment as 
function of Bjorken $x$. 
The curves (from top to bottom) correspond to deuterated butanol, ammonia and 
butanol.}
\label{fig:f}
\end{figure}

The resulting dilution factor is shown in Fig.~\ref{fig:f}
as a function of $x$ for the 3 targets used in the SMC experiment. All curves show a significant drop
at low $x$ due to the large radiative cross sections on  nuclei in this region. The rise observed
at large $x$ for ${\rm NH}_3$ and p-butanol results from the drop of the ratio $F_2^n/F_2^p$ discussed
in the previous section. \\

A further correction is needed when the diluting material itself contains some polarizable 
nuclei as for instance $^{14}N$ in the SMC ammonia target \cite{ptarg}. Several other cases
(not mentioned here) are discussed in ref. \cite{rondon}. \\

\begin{figure}[here]
\begin{center}
\mbox{\epsfig{figure=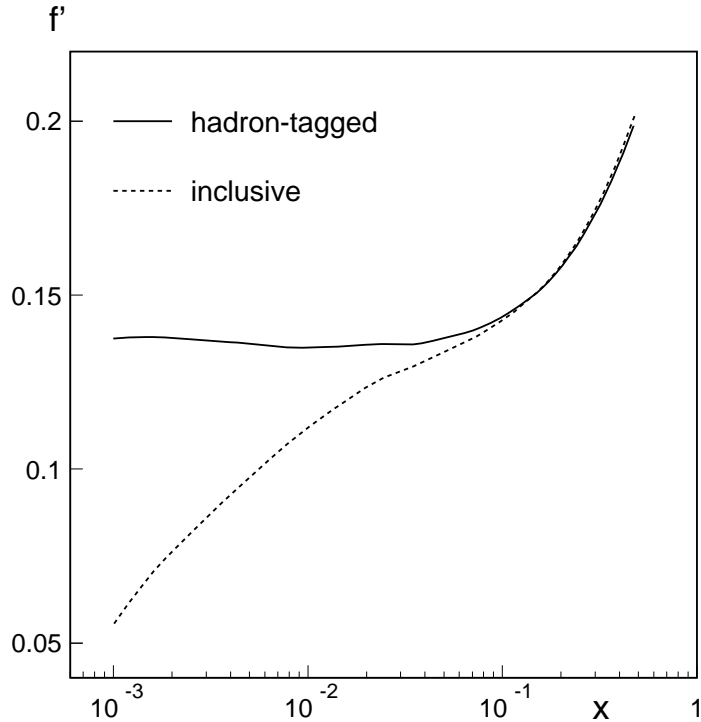,width=100mm}}
\end{center}
\caption{\small
\it
 Effective dilution factor $f'$ for hadron tagged and inclusive events
from the SMC ammonia target.}
\label{fig:dilut}
\end{figure}

In addition to dilution due to the presence of 
non-polarized material in the target, one has to take into account an additional effect
due to radiative events taking place on the polarized proton or deuteron. These radiative
contributions modify both the spin-averaged and spin dependent cross sections with respect to the value
expected from the one photon exchange process (Fig.~\ref{fig:dis}).
In general the total (one-photon
exchange $+$ radiative processes) cross sections $\sigma^T$ and the one-photon exchange $\sigma^{1 \gamma}$
cross sections are related by
\begin{equation}
\begin{array}{lcl}
{\overline \sigma}^T &=& v ~ {\overline \sigma}^{1 \gamma} ~+~ {\overline \sigma}_{tail}  \\
\Delta \sigma^T &=& v ~ \Delta \sigma^{1 \gamma} ~+~ \Delta \sigma_{tail}.
\end{array}
\end{equation}
The factor $v$ is very close to $1$ and will not be further mentioned.
The "tail" terms contain contributions from
radiative corrections to elastic scattering (with final state $\ell N \gamma$) and from the inelastic 
continuum. These contributions have been calculated in great detail 
for the spin averaged and the spin dependent cross sections and can be obtained from 
specialized computer codes \cite{polrad}. The measured asymmetry
\begin{equation}
A_{meas} = P_b ~ P_T~ f~ \frac{\Delta \sigma^T}{2 ~ {\overline \sigma^T}}
\end{equation}
can be rewritten as
\begin{equation}
A_{meas} = P_b ~ P_T ~ f ~\frac{{\overline \sigma}^{1 \gamma}}{{\overline \sigma}^T} \frac{\Delta \sigma^T}
{2 ~ {\overline \sigma ^{1 \gamma}}}
\end{equation}
where it appears that 
\begin{equation}
f' ~=~ f ~\frac{{\overline \sigma}^{1 \gamma}}{{\overline \sigma^T}}
\end{equation}
is the "effective dilution factor" taking into account radiative effects. This leads to the relation
\begin{equation}
A_{meas} ~=~ P_B ~ P_T ~ f' \Bigl ( \frac{\Delta \sigma^{1 \gamma}}{2 {\overline \sigma}^{1 \gamma}}
~+~ \frac{\Delta \sigma ^{tail}}{2 {\overline \sigma}^{1 \gamma}} \Bigr )
\end{equation}
where the asymmetry is split into 2 terms corresponding respectively to pure one-photon exchange
and to radiative effects. In the usual kinematic conditions, the second term represents a rather small
correction (less than 10 \% of the one-photon term). In contrast, the change in the dilution
factor ($f'$ compared to $f$) becomes a major effect at small $x$ : for $Q^2 \simeq 1$ GeV$^2$  and $x= 0.005$ ,
the ratio $f'/f$ is of the order of 0.66. \\
This feature makes asymmetry measurements very unaccurate at low $x$, except if a large fraction
of the radiative events can be eliminated from the data set. This can be achieved  by selecting events
with at least one high-energy hadron in the final state (e.g. by use of a calorimeter). In this
way, the radiative effects due to the elastic tail do not contribute any more and the effective
dilution factor no longer drops at low $x$. The effect of selecting hadron-tagged events
is illustrated in Fig.~\ref{fig:dilut} for the ammonia target in the SMC experiment \cite{smc_2}.

\section {Evaluation of sum rules.}
\begin{figure}[here]
\epsfxsize=9.5cm
\epsfysize=9.5cm
\hfil
\epsffile[10 20 525 550]{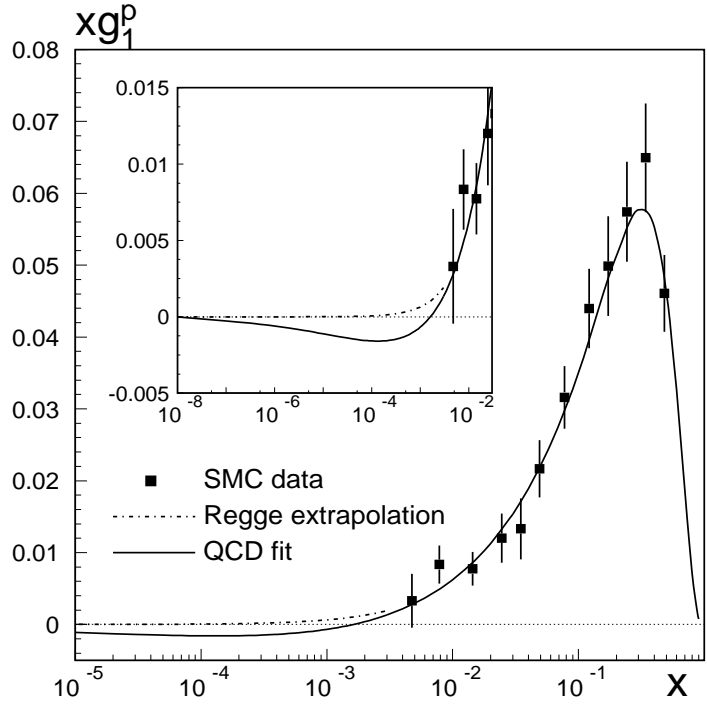}
\hfil
\caption{{\small \it {$xg_{1}^{\rm p}$ as a function of $x$; SMC data points (squares) with the total error are
shown together with the result of the QCD fit (continuous line), both at $Q^{2}=10$ GeV$^{2}$. 
For $x < 0.003$ the extrapolation assuming Regge behaviour is indicated by the dot-dashed line.
The inlet is a close-up extending to lower $x$. \cite{smcnh3} }}}
\label{fig:lowx}
\end{figure}

\vspace*{0.2cm}
In the previous sections, we have shown how the longitudinal spin structure function $g_1(x,Q^2)$
can be evaluated in the range of $x$ covered by an experiment. The evaluation of the first moment
$\Gamma_1(Q_0^2) ~=~ \int_0^1 g_1(x,Q_0^2)~dx$ involves also the unmeasured range of $x$ and is thus
not a fully measurable quantity. Splitting the integration over the measured and unmeasured range
of $x$ one obtains
\begin{equation}
\Gamma_1(Q_0^2) ~=~ \int_0^{x_{min}} g_1(x,Q_0^2)~dx + \int_{x_{min}}^{x_{max}} g_1(x,Q_0^2) ~dx
+ \int_{x_{max}}^1 g_1(x,Q_0^2) ~dx.
\end{equation}
The contribution of the last term is limited by the boundary condition $|A_1| \le 1$ and by the small value
of the unpolarized structure function $F_2$ at large $x$ and is thus not critical. In contrast, the unmeasured
contribution at low $x$ may be important and may influence significantly  sum rules involving $\Gamma_1$. \\
The value of $x_{min}$ depends on the experiment kinematics and on the choice of $Q_0^2$:
\begin{equation}
x_{min}(Q_0^2) ~=~ \frac{Q_0^2}{2~M~\nu_{max}} ~=~ \frac{Q_0^2}{2~M~y_{max}~E_{beam}}.
\end{equation}
Requiring $Q_0^2 = 1 $GeV$^2$ to remain in the DIS region and setting $y_{max} = 0.9$, we obtain
$x_{min} = 0.003$ for a beam energy of 200 GeV. This is indeed the lowest experimental value of $x$ reached
up to now
for  $Q^2 \ge 1$ GeV$^2$. The integral of $g_1(x)$ below this value can, in principle, be estimated by a
smooth extrapolation of the values obtained at higher $x$ taking into account the known asymptotic behaviour
of cross sections at high energy. An extrapolation of this kind is shown by the dot-dashed
 line in Fig.~\ref{fig:lowx}
for the SMC data on $g_1^p$.
There is however an ambiguity in this procedure since the experimental values of
$g_1$ are obtained at a different $Q^2$ for each $x$ interval and need to be evolved to a common (and arbitrary) $Q^2$ before
any extrapolation can be performed. 
For this reason, a different procedure has been adopted in the most recent evaluations of sum rules. \\
The analysis of the $Q^2$ 
dependence of $g_1(x,Q^2)$ requires  input parametrizations of the constituant spin distributions $\Delta q_i(x,Q_0^2)$
at a reference $Q^2$ value 
which is most frequently choosen to be 1 GeV$^2$. These parametrizations are then evolved according to the
evolution equations of QCD and adjusted to the experimental values of $g_1$. The curves on Fig.~\ref{fig:qcdprot} show that
they provide a satisfactory description of the data. In view of this, one may assume that the same parametrizations
remain valid below the lowest $x$ value of the measurements and use the fitted $g_1$ distribution to evaluate
the unmeasured contribution at low $x$. The continuous  line in Fig.~\ref{fig:lowx} shows that this approach leads to a very
different result  compared to the "smooth extrapolation" described before: $g_1(x)$ becomes negative below 0.001, i.e.
slightly below the lowest data point. This affects significantly the first moments of $g_1$ as well for the proton as for
the deuteron. The values obtained in the measured range and the estimated low $x$ contributions
at $Q^2 = 10$ GeV$^2$
are shown in Table 1
for the SMC experiment. \\
\vspace{5mm}
\begin{table} [here]
\begin{center}
\begin{tabular}{|c|c|c|}
\hline  \hline
  &  Measured range     &   Extrapolation  \\
  &   0.003 - 0.7       &    0.0 - 0.003   \\
\hline
$\Gamma_1^p$ &  0.131 $\pm$ 0.009  &  -0.011  \\
\hline
$\Gamma_1^d$ &  0.037 $\pm$ 0.007  &  -0.018  \\
\hline  \hline
\end{tabular}
\caption{\small
\it
Low $x$ extrapolation of $g_1^p$ and $g_1^d$  in the SMC experiment at $Q^2$ ~=~ 10 Gev$^2$ \cite{smc_2}.}
\end{center}
\end{table}

\section {Prospects and future experiments.}
\vspace*{0.2cm}

In the previous sections we have shown how the longitudinal spin structure function has been measured in
several polarized DIS experiments over the last 20 years. The precision of the results is limited  by the
statistical accuracy of the data and their range is limited by the kinematics of fixed target experiments.
There are also further limitations inherent to the physics of inclusive reactions:
the structure function  $g_1$ does not fully describe the distribution of the nucleon spin.
In particular it provides little
discrimination between quark flavors and no discrimination at all between quarks and antiquarks. \\
Additional information can be obtained by studying semi-inclusive reactions, i.e. by requiring the
presence of a given hadron in the final state. This can only be achieved in an experiment providing
particle identification. Detailed studies of semi-inclusive polarized scattering are presently under
way in the HERMES experiment.  \\
The gluon spin distribution $\Delta g$ contributes to $g_1$ only by the $Q^2$ evolution and is therefore
poorly constrained by inclusive measurements. A more direct determination of $\Delta g$ can be achieved
by selecting final states mainly poduced by the interaction of a gluon with the exchanged virtual photon.
This is the case in particular for the production of charmed quarks ($\gamma^{*} g \rightarrow c {\overline c}$),
a process presently studied by the COMPASS experiment at CERN \cite{compass}. \\
In addition to $g_1$, another structure function, the so-called transversity distribution, can be measured
in reactions where the polarization of the target nucleons is perpendicular to the beam polarization \cite{artru}. This
function is related to the transverse spin distribution of the quarks in the same way as $g_1$ is related to
their longitudinal spin distribution. Transversity will also be measured by the HERMES and COMPASS experiments. \\
The possibility to extend the kinematic domain of polarized DIS scattering has been investigated by different
study groups over the last five years. Projects have been developped for a polarized proton ring at HERA \cite{desy}
and for an electron-ion collider at BNL \cite{bnl}. In both projects, the lower limit of $x$ would be 
considerably reduced and negative values of $g_1^p$ would become observable if the presently used
extrapolation is correct. \\

\end{document}